\documentclass[preprint,showpacs,preprintnumbers,amsmath,amssymb]{revtex4}

\usepackage{graphicx}
\usepackage{dcolumn}
\usepackage{bm}

\begin{document}

\preprint{}

\title{Isospin dependence of nucleon emission and radial flow 
in heavy-ion collisions induced by high energy radioactive beams}

\author{Bao-An Li}
\email{bali@astate.edu}
\affiliation{Department of Chemistry and
Physics, P.O. Box 419, Arkansas State University, State
University, Arkansas 72467-0419, USA}
\author{Gao-Chan Yong}
\author{Wei Zuo}
\affiliation{Institute of Modern Physics, Chinese Academy of
Science, Lanzhou 730000, P.R. China} 
\affiliation{Graduate School, Chinese Academy of Science, 
Beijing 100039, P.R. China}
\date{\today}

\begin{abstract}
Using an isospin- and momentum-dependent transport model we study 
the emission of free nucleons and the nuclear radial flow  
in central heavy-ion collisions induced by high energy radioactive beams. 
The midrapidity neutron/proton ratio and its transverse momentum dependence are found very 
sensitive to the high density behavior of nuclear symmetry energy. 
The nuclear radial flow, however, depends only weakly on the symmetry energy.    
\end{abstract}
\pacs{25.70.-z, 25.70.Pq., 24.10.Lx}
\maketitle

The density dependence of nuclear symmetry energy $E_{sym}(\rho)$ 
is very important for understanding many 
interesting questions in astrophysics\cite{bethe,lat01,ibook01,bom1,steiner04}, 
the novel structures of radioactive nuclei\cite{brown,horow,furn,stone} and many 
observables of heavy-ion reactions\cite{ibook01,ireview98,dan02,dit04}. 
However, it is still very poorly known both theoretically and experimentally, especially 
at supranormal densities, see e.g., refs.\cite{bom1,zuo}. Fortunately, nuclear reactions 
induced by radioactive beams provide a unique opportunity to pin down the symmetry 
energy in a broad density range\cite{nsac02}. Moreover, they also allow us to explore the
role of isospin degree of freedom in various phenomena and the dynamics of nuclear reactions. 
In particular, central heavy-ion reactions 
induced by high energy radioactive beams up to 400 MeV/nucleon at the planned Rare Isotope 
Accelerator (RIA) will enable us to probe the high density behavior of nuclear 
symmetry energy. While a number of experimental observables constraining the $E_{sym}(\rho)$ at 
subnormal densities have been identified recently, 
see e.g., refs.\cite{ibook01,ireview98,dit04} for reviews, 
very few observables sensitive to the high density behavior of $E_{sym}(\rho)$, 
except the $\pi^-/\pi^+$ ratio\cite{li03,ligz04}, are currently known. In this work, 
we examine midrapidity nucleons and the radial flow in central heavy-ion reactions 
induced by high energy radioactive beams as potential probes of the 
symmetry energy at high densities. 

Our study is based on the isospin- and momentum-dependent transport model IBUU04\cite{lidas03}.
In this model a parameter $x$ was introduced in the single nucleon potential to mimic
various density dependences of the symmetry energy as predicted by different microscopic 
many-body theories\cite{das03}. Shown in Fig.\ 1 are examples of the symmetry energy
with the parameter $x$ going from 1 (softer) to -2 (stiffer). The recent data 
from NSCL/MSU on isospin diffusion in mid-central $^{124}Sn+^{112}Sn$ 
reactions at 50 MeV/nucleon was used to constrain the symmetry energy 
at subnormal densities\cite{betty03}. Within the IBUU04 model, a symmetry energy
$E_{sym}(\rho)\approx 31.6(\rho/\rho_0)^{1.05}$ corresponding to $x=-1$ was found to 
reproduce the isospin diffusion data very nicely\cite{chen04b}. In the present study we use
the same model and parameter set to study nucleon emissions and the radial flow for central 
$^{132}Sn+^{124}Sn$ reactions at a beam energy of 400 MeV/nucleon. 
In this reaction a compressed matter of about twice normal nuclear matter density is formed and 
lasts for about 15 fm/c\cite{ligz04}. This particular reaction system will be studied experimentally 
using the Time Projection Chamber with beams from the fast fragmentation line at RIA. 

Shown in Fig.\ 2 are the rapidity distributions of free nucleons identified as those having
local baryon densities less than $\rho_0/8$ at 30 fm/c when the dynamical freeze-out for 
particle emission is reached. It is seen that mid-rapidity nucleons are very sensitive to 
the symmetry energy. With the symmetry energy changing from being softer (x=1) to stiffer (x=-2)
more (less) neutrons (protons) are emitted at midrapidity. This is a direct consequence of the 
higher repulsive (attractive) symmetry potential for neutrons (protons) with the stiffer 
symmetry energy at supranormal densities. Moreover, neutrons appear to be more sensitive
to the variation of the symmetry energy than protons. This is because for protons the repulsive 
Coulomb potential works against the symmetry potential. It is also interesting to 
point out that the effects of the symmetry potential on nucleon emissions observed here 
are just opposite to those at intermediate energies around 50 MeV/nucleon 
where a softer symmetry energy causes more (less) emissions of 
neutrons (protons)\cite{li97}. This is because of the crossover of the 
symmetry energy functions with different density dependences 
at $\rho_0$ as shown in Fig.\ 1. Thus it is very important to carry out 
excitation function studies of particle emissions in order to map out
the symmetry energy in a broad range of density.     
  
The degree of isospin equilibrium or translucency can be measured by the rapidity distribution
of nucleon isospin asymmetry $\delta_{free}\equiv (N_n-N_p)/(N_n+N_p)$ where $N_n$ ($N_p$) 
is the multiplicity of free neutrons (protons)\cite{li04a}. Although it might be difficult to measure
directly $\delta_{free}$ because it requires the detection of neutrons, 
similar information can be extracted from ratios of light clusters,
such as, $^{3}H/^3He$, as demonstrated recently within a coalescence model\cite{chen03b,chen04a}.   
Shown in Fig.\ 3 are the rapidity distributions of $\delta_{free}$ with (upper window) 
and without (lower window) the Coulomb potential. It is interesting to see that the
$\delta_{free}$ at midrapidity is particularly sensitive to the symmetry energy.
As the parameter $x$ increases from $-2$ to $1$ the $\delta_{free}$ at midrapidity decreases
by about a factor of 2. Moreover, the forward-backward asymmetric rapidity distributions 
of $\delta_{free}$ with all four $x$ parameters indicates the apparent nuclear translucency 
during the reaction. Comparing the results shown in the upper and lower windows, 
one sees that the Coulomb potential is to reduce the value of $\delta_{free}$, while the
sensitivity to the symmetry energy remains about the same.

Concentrating on nucleons at midrapidity, we examine in Fig.\ 4 the differential 
neutron/proton ratio $dN_n/dN_p$ as a function of transverse momentum $p_t$. The solid line in 
the figure is the average $(n/p)_{sys}$ ratio of the reaction system. In the low (high) 
$p_t$ part the $dN_n/dN_p$ is significantly higher (lower) than the $(n/p)_{sys}$. 
Moreover, the low $p_t$ part of the $dN_n/dN_p$ ratio is more sensitive to the 
symmetry energy than the high $p_t$ part. Both observations are due to the momentum
dependence of the symmetry potential. As shown earlier by one of us the symmetry 
potential decreases with increasing nucleon momentum in agreement with the Lane 
potential extracted from nucleon-nucleus scattering data\cite{lidas03,li04}.    
Of course, the symmetry potential is also density dependent. Generally, high $p_t$
particles are more likely to come from high density regions where the symmetry potential is 
stronger. However, if effects of the momentum dependence are stronger than those of the 
density dependence, the low $p_t$ particles are then more sensitive to the symmetry potential
as observed here. 

It is also seen from Fig.\ 4 that the $dN_n/dN_p$ ratios with different 
$x$ parameters are almost parallel with each other except in the high $p_t$ region. 
It is well known within the blast wave picture of central havey-ion reactions that 
the inverse slope of the particle spectra, i.e., the apparent temperature, reflects the 
combined effects of the true temperature and the radial flow 
velocity of the compressed matter in the exploding fireball\cite{siemens}.
Our observation here about the slopes of the $dN_n/dN_p$ ratios 
indicates that the symmetry energy has a weak influence on
the overall thermalization and the radial flow. To be more specific we study in the following the
isospin dependence of radial flow. It is worth noting that to our best knowledge 
this kind of study has never been carried out before. In anticipation of the 
coming experiments at RIA, it is important to
identify what role the isospin degree of freedom may play in the radial flow which is 
in its own right an interesting phenomenon. We extract the average radial flow velocity 
at radius $r$ from 
\begin{equation}\label{beta}
\beta(r)=\frac{1}{N(r)}\sum_{i}^{N}\vec{p}_i\cdot \vec{r}_i/E_i,
\end{equation}
where $N(r)$ is the number of particles including bound ones in the spherical shell between 
radius $r$ and $r+dr$, $\vec{p}_i, \vec{r}_i$ and $E_i$ are the momentum, coordinate 
and energy of the particle $i$. 
As an example, we show in Fig\ 5 the radial flow velocities of neutrons and protons and their 
densities as functions of radius $r$ measured from the center of mass of the reaction with $x=-1$.
It is seen that the radial flow velocity increases with the radius $r$ similar to the Hubble 
expansion of the universe. This observation is consistent with other studies on the 
radial flow in heavy-ion reactions, see, e.g., refs.\cite{liko95,liko96}. 
Protons, especially at larger radii, are flowing 
with a slightly higher velocity due to the Coulomb repulsion, although they have about the 
same density profile as neutrons. To explore effects of the symmetry energy on the radial flow    
we show in Fig.\ 6 the position-averaged nucleon radial flow velocity $<\beta>$ with and without the 
Coulomb potential as a function of the parameter $x$. The value of $<\beta>$ can be readily 
extracted experimentally from the particle spectra using the Siemens-Rasmussen formula\cite{siemens}.  
In our calculations it is obtained from eq.\ref{beta} by setting $r=0$ and $dr=\infty$.
Formally, $<\beta>$ is related to $\beta(r)$ via
\begin{equation}
<\beta>=\frac{1}{N_{total}}\int^{\infty}_{0}\beta(r)\rho(r)4\pi r^2 dr,
\end{equation} 
where $N_{total}$ is the total number of neutrons or protons. 
Without the Coulomb potential, the radial flow of neutrons is faster than that of protons 
because of the repulsive (attractive)
symmetry potential for neutrons (protons). The difference in $<\beta>$ for neutrons and protons
is the largest for the stiffest symmetry energy considered here as one expects. As the symmetry
energy becomes softer the difference disappears gradually. However, the overall effect of the symmetry
energy on the radial flow is small, even for the stiffest symmetry energy with $x=-2$ the 
effect is only about 4\%. This is because
the pressure of the fireball is dominated by the kinetic contribution. Moreover, the compressional 
contribution to the pressure is overwhelmingly dominated by the isoscalar interactions. 
For protons, however, as the Coulomb potential is turned on, 
because of its dominance over the symmetry potential the whole picture changes.
In fact, the Coulomb potential almost cancels out the effect of the symmetry 
potential at $x=-2$. As the symmetry energy becomes softer, the radial flow for protons becomes
higher than that for neutrons. With the softest symmetry energy considered here, i.e, with the 
parameter $x=1$, the Coulomb potential results in a higher radial flow velocity 
for protons than neutrons by about 3\%. The radial flow for neutrons is 
only weakly affected through a secondary effect because the reaction dynamics 
is slightly modified by the Coulomb interaction. 

In summary, using an isospin- and momentum-dependent transport model
we have studied nucleon emissions and the nuclear radial flow as potential 
probes of the symmetry energy at high densities. 
We found that the isospin asymmetry of midrapidity nucleons and its 
transverse momentum dependence are very sensitive to the high density behavior 
of nuclear symmetry energy. These observables together with the $\pi^-/\pi^+$ 
ratio reported earlier will by very useful for studying the equation of state of dense 
neutron-rich matter at RIA. We have also studied the isospin dependence of 
nuclear radial flow. Our findings clearly indicate that the 
isospin degree of freedom plays an important role in the nuclear radial flow through the
competition between the Coulomb and the symmetry potential. However, the net effect of the 
symmetry energy on the radial flow is small.  

The work of B.A. Li is supported in part by the National Science Foundation of the
United States under grant No. PHYS-0243571 and PHYS0354572, and the NASA-Arkansas 
Space Grants Consortium. The work of G.C. Yong and W. Zuo is supported in part by 
the Chinese Academy of Science Knowledge Innovation Project (KECK2-SW-N02),
Major State Basic Research Development Program (G2000077400), the
National Natural Science Foundation of China (10235030) and the
Important Pare-Research Project (2002CAB00200) of the Chinese
Ministry of Science and Technology.

\newpage
\begin{figure}
\includegraphics[scale=0.55,angle=-90]{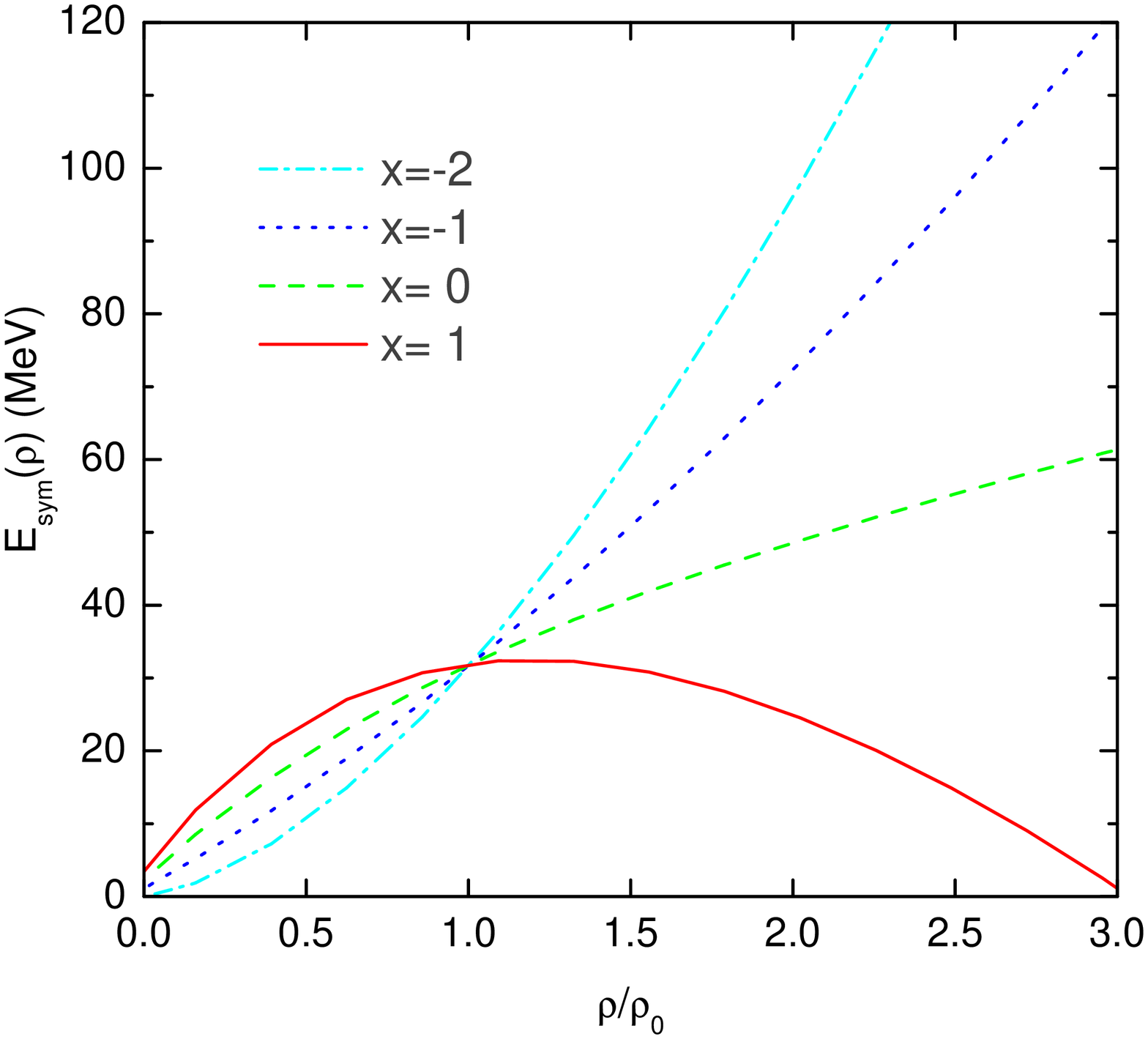}
\vspace{1.cm} \caption{{\protect\small (color online) Nuclear symmetry
energy as a function of density with different $x$ parameter.}}
\label{esym}
\end{figure}

\begin{figure}[h]
  \begin{center}
    \rotatebox{270}{\includegraphics*[width=0.6\textwidth]{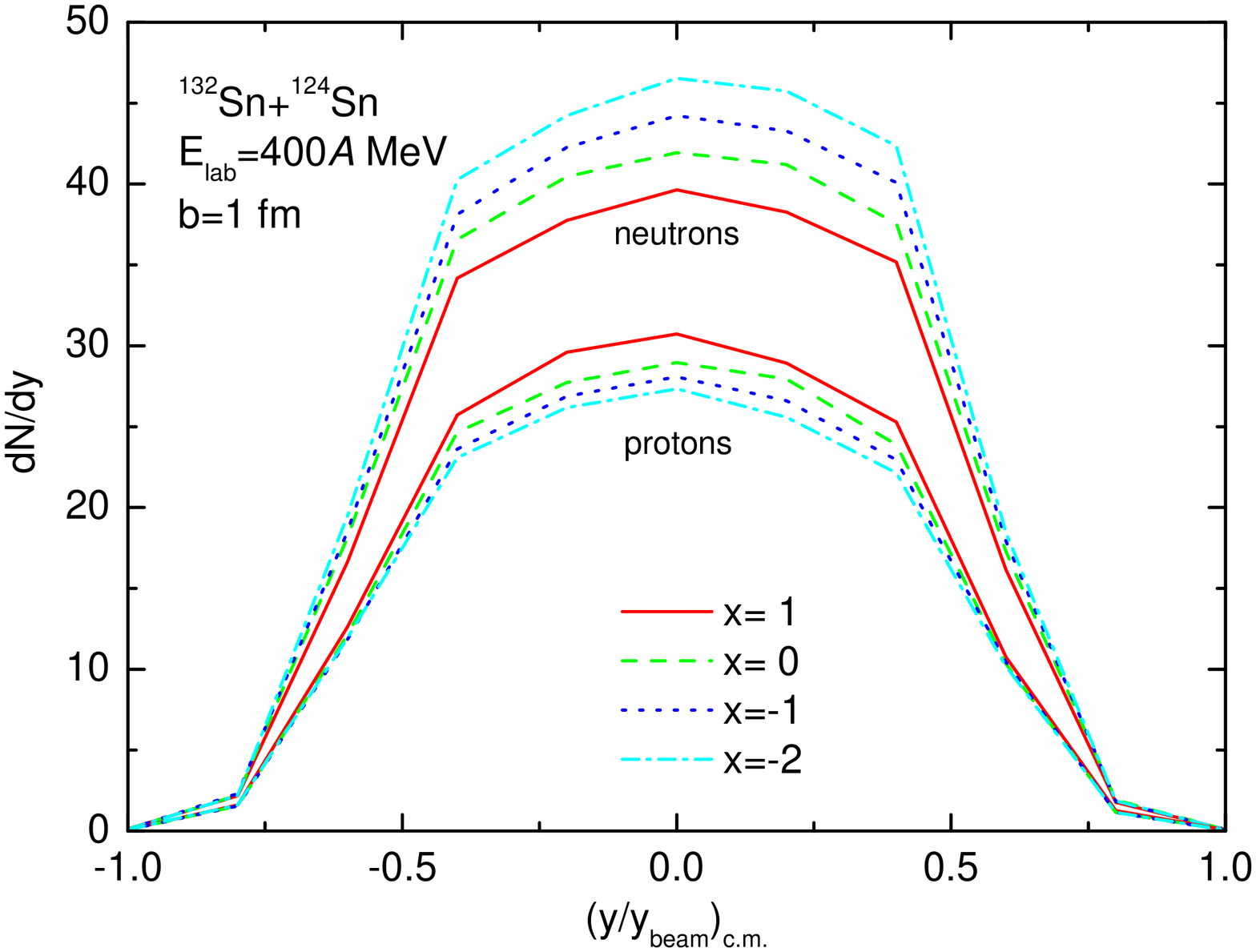}}
    \caption{(color online) Rapidity distributions of free neutrons and protons in
    the reaction of $^{132}Sn+^{124}Sn$ at $E_{beam}$/A = 400 MeV and an impact 
parameter of 1 fm.}
  \end{center}
\end{figure}
\begin{figure}[h]
  \begin{center}
    \rotatebox{270}{\includegraphics*[width=0.5\textwidth]{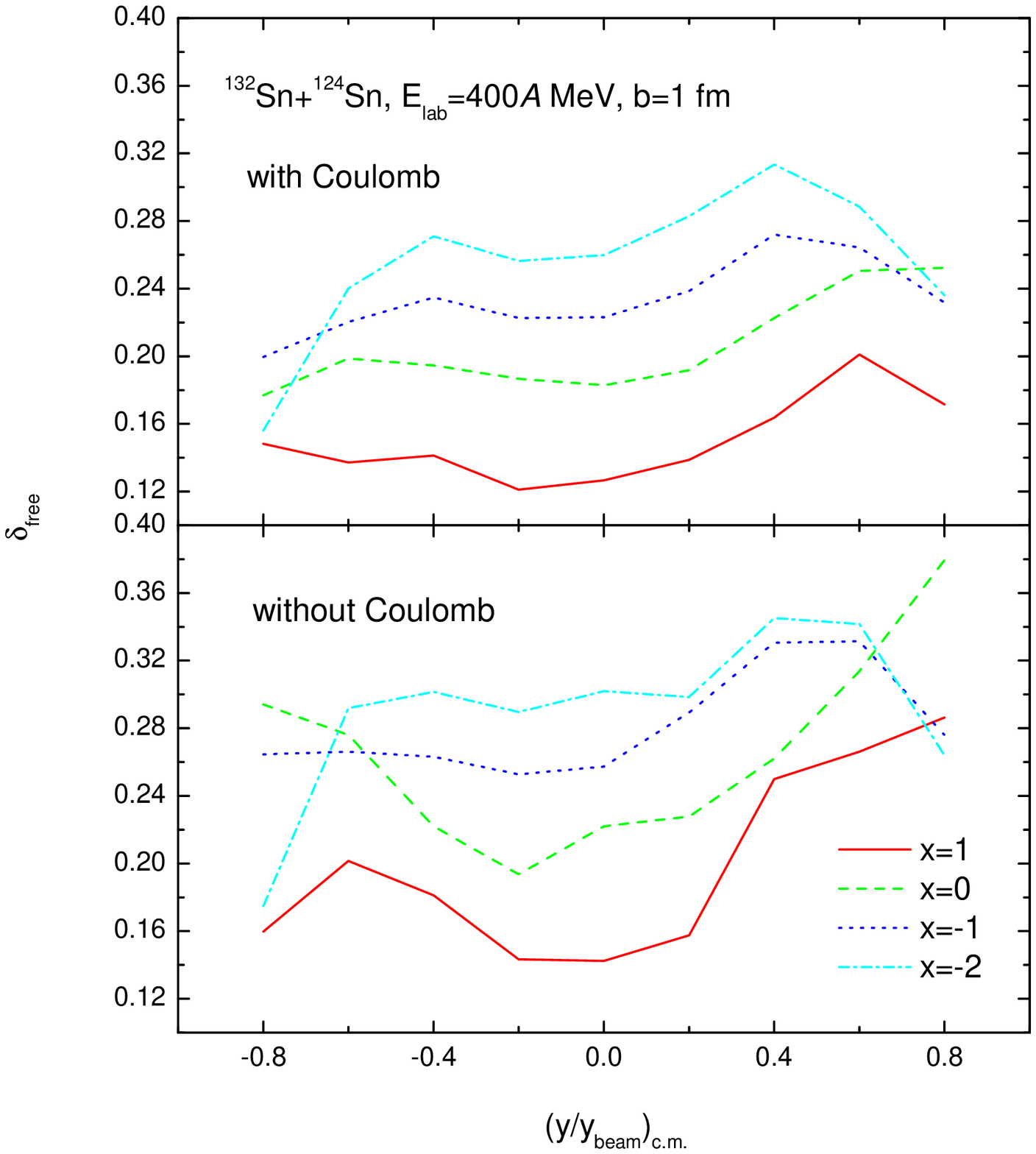}}
    \caption{(color online) Rapidity distribution of isospin asymmetry of free
    nucleons in the same reaction as in Fig.\ 2 
    with (upper window) and without (lower window) the Coulomb potential.}
  \end{center}
\end{figure}
\begin{figure}[h]
  \begin{center}
    \rotatebox{270}{\includegraphics*[width=0.5\textwidth]{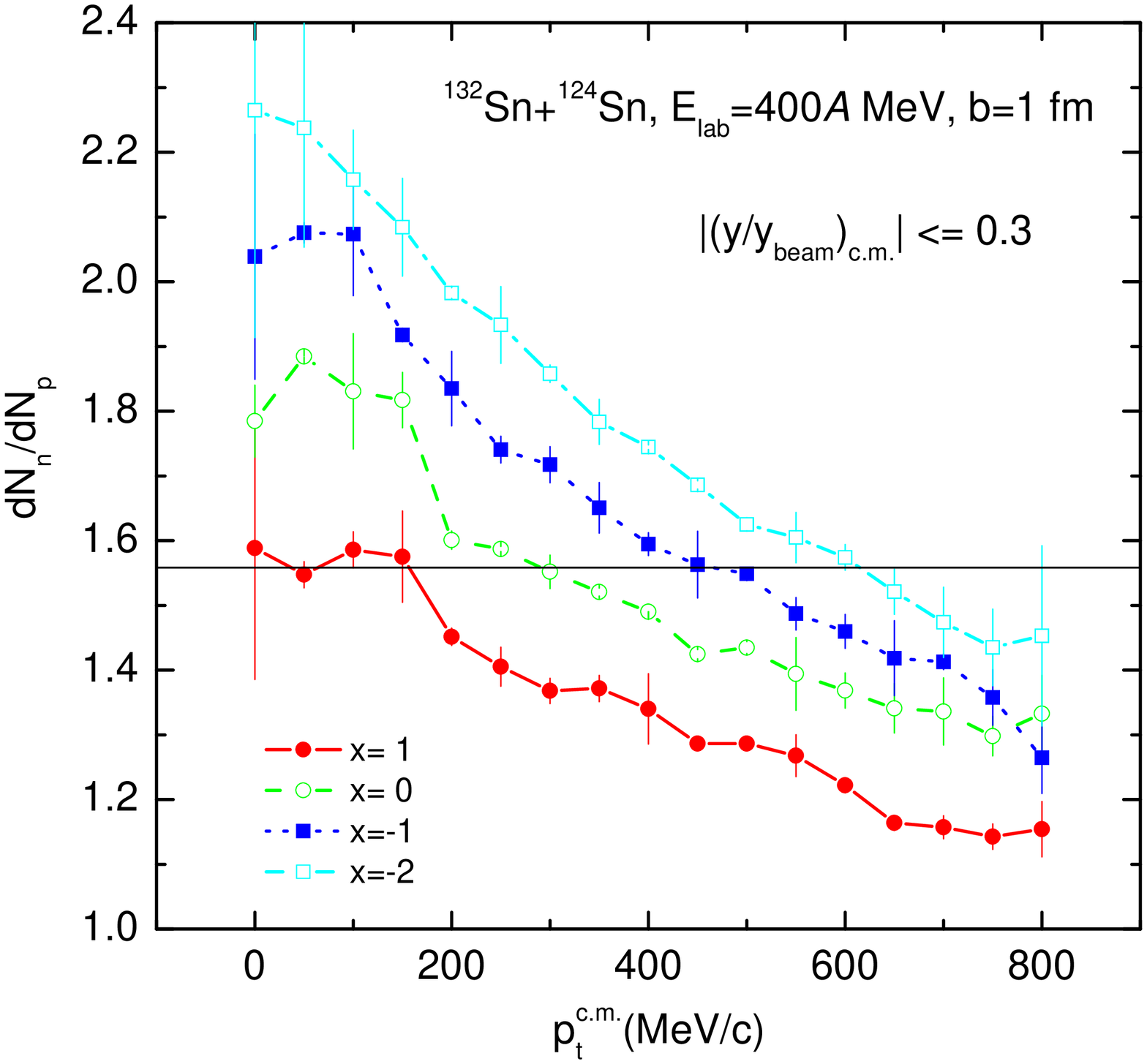}}
    \caption{(color online) Transverse momentum dependence of the differential 
              neutron/proton ratio at mid-rapidity in the same reaction 
              as in Fig.\ 2.}
  \end{center}
\end{figure}
\begin{figure}[h]
  \begin{center}
    \rotatebox{270}{\includegraphics*[width=0.5\textwidth]{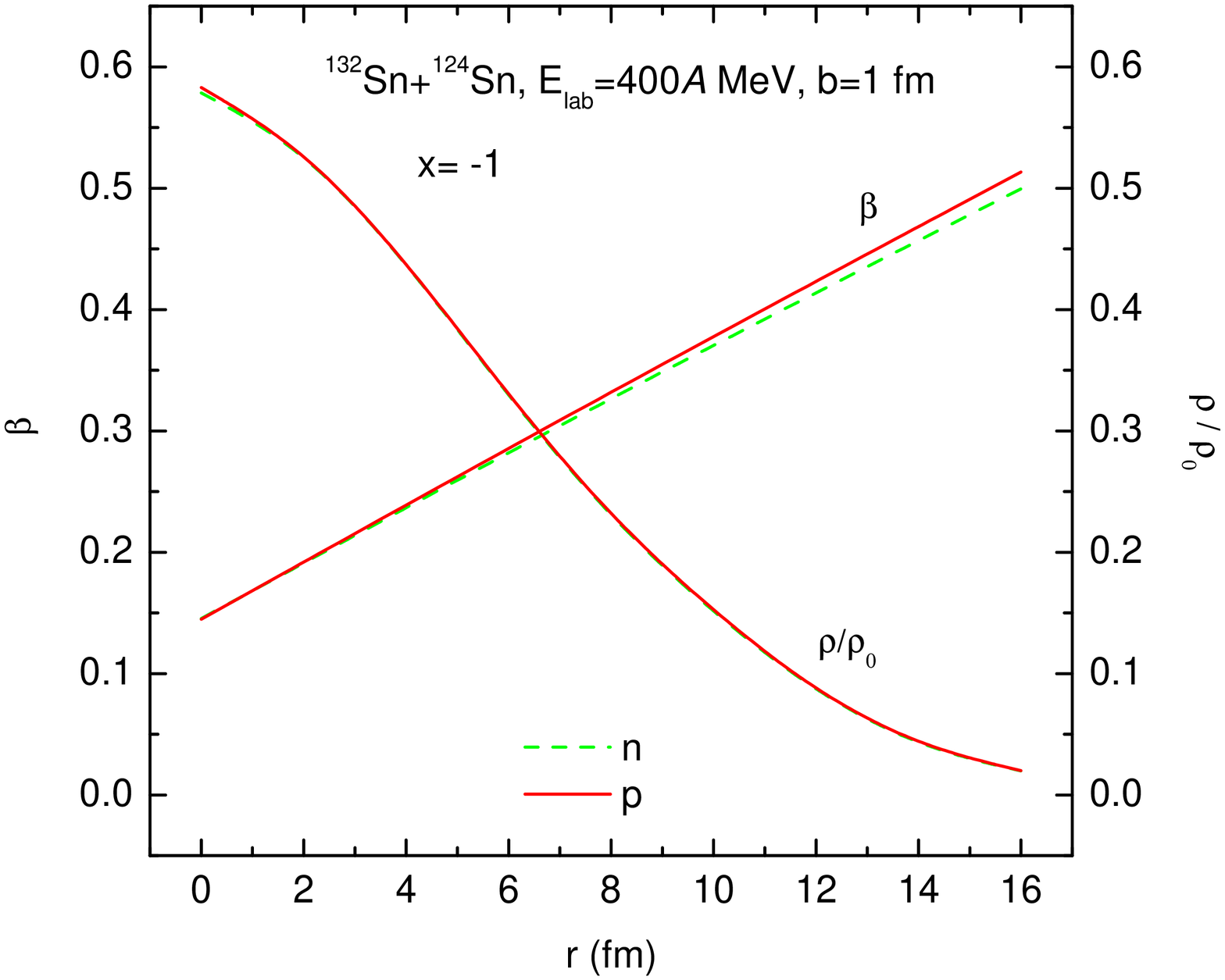}}
    \caption{(color online) Nuclear radial flow velocity and density as a function of position 
             for the same reaction as in Fig.\ 2.}
  \end{center}
\end{figure}
\begin{figure}[h]
  \begin{center}
    \rotatebox{270}{\includegraphics*[width=0.5\textwidth]{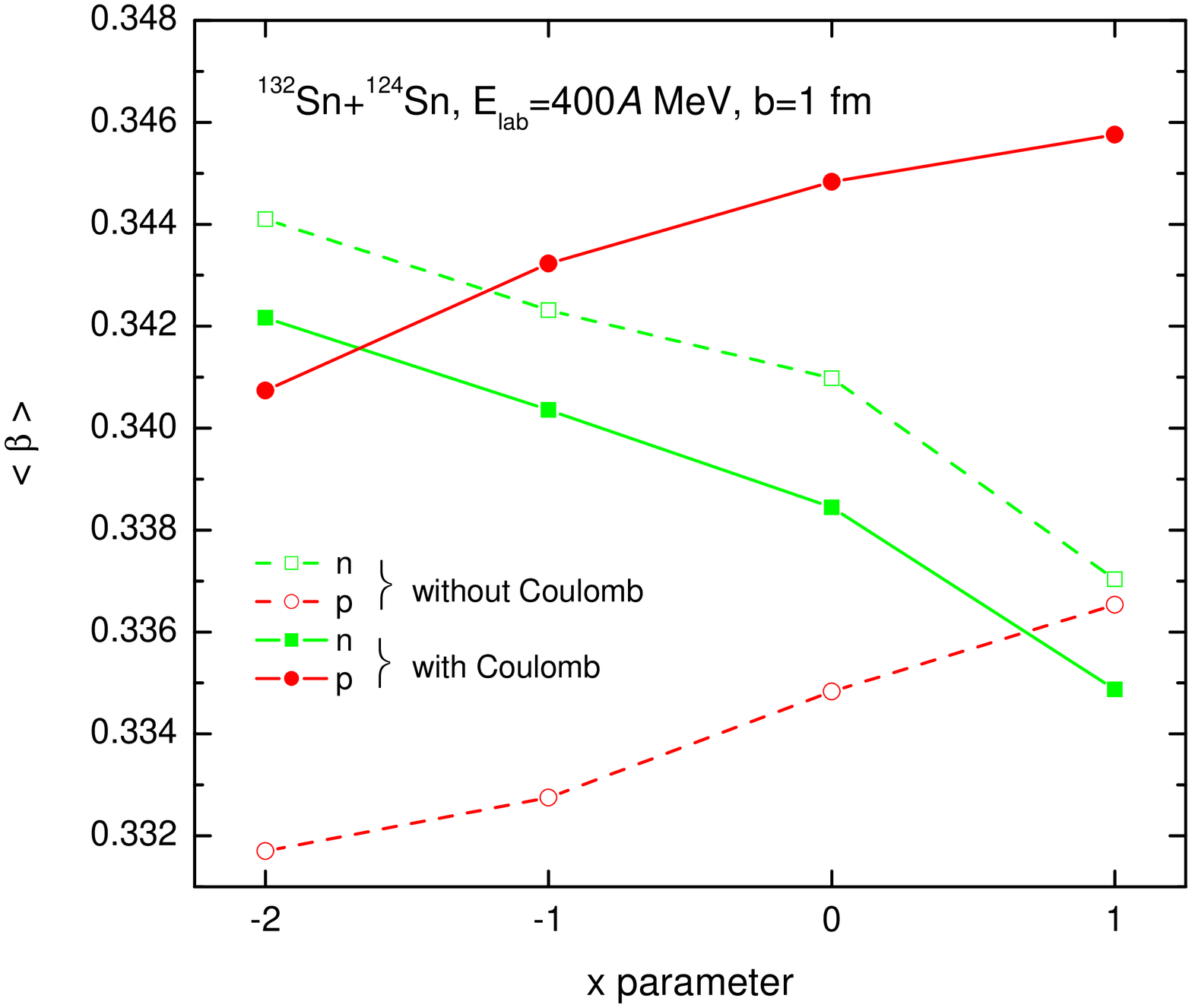}}
    \caption{(color online) Symmetry energy dependence of the position-averaged 
             radial flow velocity for the same reaction as in Fig.2.}
   \end{center}
\end{figure}

\end{document}